\documentstyle[sprocl]{article}

\input{psfig}

\def\be{\begin{equation}}
\def\ee{\end{equation}}
\def\bea{\begin{eqnarray}}
\def\eea{\end{eqnarray}}

\newcommand{\1}{1\!\!\!\bot}
\newcommand{\Tr}{\mathop{\rm Tr}\nolimits}
\def\x{{\bf x}}
\def\k{{\bf k}}
\def\emu{{\bf e_{\mu}}}
\def\ed{{\bf e}_d}
\def\vsig{\vec{\mbox{$\sigma$}}}
\def\vca{\vec{\mbox{$\cal A$}}}

\begin{document}

\title{
 GAUGE FIXING AND GLUON PROPAGATOR
        IN $\lambda$-GAUGES~\footnote{Poster presented by
A. Cucchieri
at the {\it Strong and Electroweak Matter `98} Conference in
Copenhagen, December 2-5, 1998.}
}

\author{A. CUCCHIERI and T. MENDES}

\address{Universit\"at Bielefeld, Fakult\"at f\"ur Physik\\
         D-33615 Bielefeld, GERMANY\\ 
         E-mail: cucchieri@physik.uni-bielefeld.de,
                 mendes@physik.uni-bielefeld.de}


\maketitle\abstracts{ 
We discuss critical slowing-down of several gauge-fixing 
algorithms for the so-called $\lambda$-gauges in the SU(2) case at zero 
temperature. For these gauges we also
evaluate the gluon propagator using different 
definitions of the lattice gluon field, corresponding to discretization 
errors of different orders.
}

\section{Gauge Fixing}
     Efficiency of gauge fixing algorithms is an important
     issue in the study of gauge-dependent quantities such as
     the gluon propagator on the lattice. We 
     study critical slowing-down (CSD) by determining
     the dynamic critical exponent $z$ for various
     algorithms, using optimal gauge-fixing quantities. 

     We consider the $SU(2)$ case of the standard Wilson
     action and, in order to impose the $\lambda$-gauge condition
     \cite{A4},
     we look for a local minimum of the functional
\vspace{-2mm}
\begin{equation}
       {\cal E}_{\lambda}[g]\,\equiv\, -\sum_{\x}
   \Tr\Bigl[\lambda \, g(\x)U_{d}(\x)g^{\dagger}(\x+\ed ) \,+\,
   \sum_{\mu=1}^{d-1} g(\x)U_{\mu}(\x)g^{\dagger}(\x+\emu)\Bigr]
\;.
\vspace{-1mm}
\end{equation}
This corresponds to the condition
\vspace{-2mm}
\begin{equation}
      \lambda\, \partial_d A_d^{(g)}(\x) \,+\,
       \sum_{\mu=1}^{d-1} \partial_{\mu} A_{\mu}^{(g)}(\x) \;=\; 0\;,
\vspace{-1mm}
\end{equation}
where $A$ is the gluon field $\,(U-U^{\dagger})/2i\,$.
The case $\lambda=1$ is the Landau gauge, whereas
$\lambda=0$ is the Coulomb gauge.
     At each step of a gauge-fixing algorithm, $g(\x)$ is
     updated to a new value $g^{(new)}(\x)$ in order to minimize
     ${\cal E}[g]$.
     We consider the following algorithms: \cite{A1,A1b,A2}
     the {\em local} methods of
     Los Alamos (exact local minimization), 
     Cornell (where the functional
     is minimized in the direction of the local downhill gradient
     of ${\cal E}[g]$), overrelaxation
     (where local minimization is combined with an
     ``energy-preserving'' update by means of a tuning parameter
     $\omega$), and
     stochastic overrelaxation (where either type of update
     is randomly chosen at each step), as well as the {\em global}
     method of Fourier acceleration. This algorithm can be seen
     as a generalization of the Cornell method, whose update
     is given by
\be
      g^{(new)}(\x)\;\propto\; \left[ \1 -
         \alpha \,(\nabla \cdot A^{(g)})(\x) \right]\,g(\x)\;.
\ee
     In fact, for the Fourier acceleration method we have \cite{A1b,A2}
\be
\vspace{-2mm}
      g^{(new)}(\x)\;\propto\; 
      \left[ \1 - \alpha\,p^2_{max}\,\Delta^{-1}_{\lambda} 
         (\nabla \cdot A^{(g)})(\x)\right] \,g(\x) 
      \;.
\vspace{1mm}
\ee
     The inverse of the $\lambda$-Laplacian
$ \Delta_{\lambda} \equiv \lambda \partial_d^2 + \sum_{\mu=1}^{d-1} 
\partial_{\mu}^2 $
     is usually obtained from $\,{\widehat F}^{-1}
     \,p^{-2}_{\lambda}(\k)\,{\widehat F}\,$,
     where $p^{2}_{\lambda}(\k)$ are the eingevalues
     of $\Delta_{\lambda}$,
     and ${\widehat F}$ indicates the Fourier transform.
     In the Landau case we have introduced \cite{A1b} a
multigrid implementation 
     of this method (where the inverse Laplacian is computed using a
     multigrid algorithm), showing improved behavior and 
     applicability.

     In order to monitor the convergence of these algorithms,
     we evaluate at each step the quantity 
     $|\nabla \cdot A^{(g)}|^2$,
     as well as the quantity
\vspace{-1mm}
\begin{equation}
     Q\;\propto\; \sum_{\nu} \sum_{c} \sum_{x_{\nu}}
           \, [Q_{\nu}^{c}(x_{\nu})\,-\,{\overline Q}_{\nu}^{c}]^2 \,/ \,
                   [{\overline Q}_{\nu}^{c}]^2
\vspace{-1mm}
\ee 
  where ${\overline Q}_{\nu}^{c} \equiv 
       (1/N_{\nu})\sum_{x_{\nu}} Q_{\nu}^{c}(x_{\nu})$
      and $Q_{\nu}^{c}(x_{\nu})\equiv 
       \sum_{\mu\neq\nu} \sum_{x_{\mu}} (A^{(g)})_{\nu}^{c}(\x) \,$,
     which is seen to be 
     particularly sensitive to the goodness of the gauge fixing \cite{A1}.
     For these two quantities we expect to observe an
     exponential decay with the number of iterations $t$. In the limit 
     of large $t$ we can introduce a relaxation time $\tau$
     such that the decay is given by 
     $\,\exp(-t/\tau)\,$. As the lattice size $N$ is
     increased, and at constant physics (see Section \ref{4d}), $\tau$ will 
     grow as $\tau \sim N^z\,$, where $z$ is the 
     dynamic critical exponent of the algorithm. 
     [Note that all the algorithms we consider, except for
     Los Alamos, need tuning.]

     In the Landau case
     we have simulated \cite{A1,A1b,A2} the
     above algorithms in the 2d case at finite values of the coupling
     $\beta$, and in the 4d case at $\beta = \infty$. 
     We have found that, as expected, the Los Alamos algorithm has $z\sim 2$,
     the improved local algorithms (i.e.\ Cornell,
     overrelaxation and stochastic overrelaxation) have $z\sim 1$,
     and that $z\sim 0$ for Fourier acceleration.
     These results were also verified analytically \cite{A2}
     at  $\beta = \infty$. The multigrid implementation of Fourier
     acceleration is found to be considerably more stable and easier to
     tune than the standard method, and it can be used efficiently on 
     parallel machines.
     Here we extend this study to the 4d case at finite $\beta$
and to values of $\lambda$ different from 1.

\subsection{Simulations and Results}
\label{4d}

In order to analyze CSD for an algorithm, we have to evaluate $\tau$
for different combinations of lattice size $N$ and coupling $\beta$, but at
constant physics. This is done easily in $d=2$ by keeping the 
product $N \sqrt\sigma$ constant, since the string tension $\sigma$ is 
a known function of $\beta$. No such function is available in $d=4$.
In this case we use the constant physics obtained \cite{A3}
by keeping a lattice definition of the running coupling $\alpha_s$
constant.
This prescription is well described by the fit $ \beta = 1.905+0.308\ln N $
for $\beta \geq 8$. In particular, we consider $N = 8$ at $\beta = 2.6957$,
$N = 12$ at $\beta = 2.8485$ and $N = 16$ at $\beta = 2.9586$.
By fitting our data for the the relaxation time $\tau$ to
the function $\tau = c N^z$ we
obtain the dynamic critical exponents $z$ for the various
algorithms (neglecting finite-size effects). Our
previous results \cite{A1,A1b,A2} are confirmed also
in the four-dimensional case at finite values of $\beta$:
the Los Alamos method has $z\sim 2$, while the
improved local algorithms have $z\sim 1$ and the Fourier
acceleration method shows $z\sim 0$.

All these results were obtained using the stopping criterion
$\,|\nabla \cdot A|^2\leq 10^{-12}\,$. We also check how the
performance of the various algorithms is modified if the
condition $ Q \leq 10^{-12}\,$ is used. 
For the case $N = 16$ and $\beta = 2.9586$ we have found,
in agreement with our previous findings \cite{A1,A1b,A2}, that 
for the stochastic overrelaxation and Fourier acceleration methods
the number of sweeps is essentially independent of the quantity 
chosen for the stopping criterion, while the Los Alamos,
Cornell and overrelaxation methods require about $20 \%$ more sweeps
if the quantity $ Q $ is considered.

Finally, in Table \ref{tab4} we report 
the number of sweeps needed to satisfy the condition 
$\,|\nabla \cdot A|^2\leq 10^{-12}\,$, obtained for 
lattice size $N=8$ and coupling $\beta=2.6957$ 
for the various algorithms for two different values of
$\lambda$, namely $ 1 $ and $ 0.5 $.
\small \tabcolsep0.5ex
\begin{table}
\caption{Average number of iterations for the various gauge-fixing
         algorithms and for different $\lambda$-gauges.
         We consider $N = 8$ at $\beta=2.6957$.}
\begin{center}
          \begin{tabular}[t]{|c|c|c|c|c|c|c|}
            \hline
$\lambda$ & LOS & COR & OVE & STO & FFTFA & MGFA \\
            \hline
 1 & 631(42) & 155(11) & 137(12) & 159(8) & 95(11) & 94(10) \\
            \hline
0.5 & 772(44) & 176(21) & 143(9) & 178(10) &  106(15) & 93(6) \\
            \hline
          \end{tabular}
\end{center}
\label{tab4}
\end{table}
\normalsize
From these data we see that the performance of the algorithms
seems to depend weakly on the value of $\lambda$.


\section{Gluon Propagator}

    The study of the infrared behavior of the gluon
    propagator at zero temperature~\cite{B1} provides a powerful tool 
    for gaining insight into the physics of confinement in non-Abelian
    gauge theories. In high-temperature QCD,
    the long-distance behavior of the gluon propagator is directly
    related to the electric and magnetic screening lengths \cite{B2,B3}.
    Here we study the gluon propagator at zero temperature,
    using different definitions of the lattice gluon field, and
    considering different $\lambda$-gauges. To this end, let
    us define the gluon propagator in
    momentum space as
\vspace{-2mm}
\begin{eqnarray}
    D(k) &\equiv& \frac{1}{9\,V} \sum_{\mu=1}^{d-1} D_{\mu}(k) 
    \nonumber \\
    D_{\mu}(k) &\equiv& 
    \sum_c \, \Bigl[ \sum_t
      \cos (2 \pi k t)\, Q_{\mu}^c(t) \Bigr]^2 \,+\,
     \Bigl[ \sum_t \sin (2 \pi k t)\, Q_{\mu}^c(t) \Bigr]^2
\vspace{-3mm}
\end{eqnarray}
    where $Q_{\mu}^c(t) \;\equiv\; \sum_{x,y,z} A_{\mu}^c(x,y,z,t)$
    and we set ${\bf k}=(0,0,0,k)$.
    For the lattice gluon field $A_{\mu}$ we consider several possible
    definitions, leading to discretization errors of different orders.
    For example, we define:
\vspace{-1mm}
\begin{equation}
    A_{\mu}^{(1)}(\x) \;\equiv\;
     \frac{U_{\mu}(\x)\,-\,U_{\mu}^{\dagger}(\x)}{2i}
   \qquad \mbox{and} \quad
     A_{\mu}^{(2)}(\x) \;\equiv\;
     \frac{[U_{\mu}(\x)]^2\,-\,[U_{\mu}^{\dagger}(\x)]^2}{4i}\;.
\end{equation}
    If we set $U_{\mu}(\x)\equiv \exp[i a g_0 \,\vsig \cdot \vca(\x)]$,
   we obtain that both $A_{\mu}^{(1)}(\x)$ and
    $A_{\mu}^{(2)}(\x)$
    are equal to $a g_0 \,\vsig \cdot \vca(\x)$ plus
    terms of order $a^3\,g_0^3$. We can also consider
\begin{equation}
\vspace{-2mm}
    A_{\mu}^{(3)}(\x) \;\equiv\;
    4/3\,A_{\mu}^{(1)}(\x) \,-\,
   1/3\,A_{\mu}^{(2)}(\x) 
\vspace{-1mm}
\end{equation}
    and
\vspace{-2mm}
\be
    A_{\mu}^{(4)}(\x) \;\equiv\;
    64/45\,A_{\mu}^{(1)}(\x) \,-\,
   20/45\,A_{\mu}^{(2)}(\x) \,+\,
    1/45\,\frac{[U_{\mu}(\x)]^4\,-\,[U_{\mu}^{\dagger}(\x)]^4}{8i}\;.
\vspace{-1mm}
\ee
It is easy to check that $A_{\mu}^{(3)}(\x)=a g_0 \,\vsig
\cdot
\vca(\x)$ plus terms of order $a^5\,g_0^5$, and that
$A_{\mu}^{(4)}(\x)=a g_0 \,\vsig \cdot
\vca(\x)$ plus terms of order $a^7\,g_0^7$. 
The definitions $A^{(1)}$ and $A^{(2)}$ were recently
considered by Giusti et al \cite{B4}. They find that the
corresponding gluon propagators are equal modulo a
constant factor. We perform a similar study here,
evaluating $D^{(i)}(k)$ using the different definitions
of the gluon field $A_{\mu}^{(i)}$ given above, and we try
to give an interpretation to this constant factor.

Notice that $D_{1}^{(i)}(k) \,+\, D_{2}^{(i)}(k)$ is
related to the gluon propagator used by Karsch
et al \cite{B2,B3} for the evaluation of the magnetic screening mass,
and that this screening mass is invariant under rescaling
of the propagators by a constant factor.

\subsection{Simulations and Results}

We have performed simulations at several values of $\beta$,
for lattice volumes $V=8^4$,$12^4$, 
and for the gauge parameter $\lambda=1$ (Landau gauge) and
$\lambda=0.5$. We obtain that, in all cases, the
four propagators $D^{(i)}(k)$ are equal modulo a
constant factor.
Let us notice that this proportionality constant 
between different discretizations of the
gluon propagator may be explained [at least for the simple cases
$D^{(1)}(k)$ and $D^{(2)}(k)$] as a tadpole renormalization \cite{B5}. 
In fact, let us consider the
tadpole-improved link ${\widetilde U}_{\mu}(\x)\equiv U_{\mu}(\x)/u_0$,
where $u_0$ is the mean link in Landau gauge.
Then $D^{(1)}(k)$ gets multiplied by a factor
$u_0^{-2}$, and $D^{(2)}(k)$ by $u_0^{-4}$.
At $\beta=2.2$ and $\lambda=1$ we have $u_0^{2}= 0.6790(1)$. Thus,
using tadpole-improved operators, 
the discrepancy $D^{(1)}(k)/D^{(2)}(k)$
is reduced from 1.862(9) to 1.264(6).
(A similar analysis can be done at $\lambda=0.5$.)

The gluon propagator is a gauge-dependent quantity. At finite
temperature, however, the pole masses obtained from the exponential
decay of the gluon correlation functions at large spatial separations
were proven to be gauge-independent, and therefore
gluon propagators in different gauges should be related by
a constant factor, as found in \cite{B3}.
Here we consider only the zero-temperature case, and we
expect to observe gauge dependence. We have checked that
this is indeed the case, i.e.\ we observe gauge dependence
for each given discretization $D^{(i)}$. Nevertheless, we see that,
within a gauge, the behavior of the gluon propagator is independent
of the discretization modulo a renormalization factor. We plan to extend
this study to the case of QCD at high temperature. A similar behavior
in that case would mean that the screening masses are independent of the
definition used for the lattice gluon field.

\section*{Acknowledgments}
This work was partially supported by the TMR network
Finite Temperature Phase Transitions in Particle Physics,
EU contract no.: ERBFMRX-CT97-0122.

\end{document}